\documentclass{ws-p10x7}

\begin{document}

\title{Photon Structure}

\author{Stefan S\"oldner-Rembold}

\address{CERN, CH-1211 Geneva 23, Switzerland
\\E-mail: stefan.soldner-rembold@cern.ch}

\twocolumn[\maketitle\abstract{
The LEP experiments measure the QED and QCD structure of the
photon in deep-inelastic electron-photon scattering. The status
of these measurements is discussed in this short review.
}]
\section{Kinematics}
At LEP the virtuality of the ``probing'' photon is
$Q^2=-q^2$ (the negative squared four-momentum of the photon) 
and the virtuality of the ``probed'' photon is $P^2=-p^2\approx 0$. 
The deep-inelastic scattering cross-section is written as
 \begin{eqnarray}
\lefteqn{\frac{{\rm d}^2\sigma_{\rm e\gamma\rightarrow {\rm e+hadrons}}}{{\rm d}
x{\rm d}Q^2}=
\frac{2\pi\alpha^2}{x\,Q^{4}}}\\ & & 
  \left[ \left( 1+(1-y)^2\right) F_2^{\gamma}(x,Q^2) - y^{2}
F_{\rm L}^{\gamma}(x,Q^2)\right],\nonumber
\label{eq-eq1}
 \end{eqnarray}
where $\alpha$ is the fine structure constant, $x$ and $y$ are
are the usual dimensionless variables of deep-inelastic scattering and
$W^2=(q+p)^2$ is the squared invariant mass of the hadronic final state.
The scaling variable $x$ is given by 
\begin{equation}
x=\frac{Q^2}{Q^2+W^2+P^2}.
\label{eq2}
\end{equation}
The term proportional to $F_{\rm L}^{\gamma}(x,Q^2)$ is small
and is therefore usually neglected. In leading order the structure function 
$F_2^{\gamma}(x,Q^2)$ can be identified with the sum over the parton
densities of the photon weighted by the square of the parton's charge.
\section{QED Structure Functions}
\begin{figure}[hbtp]
\begin{center}
     \mbox{
         \epsfxsize=175pt
          \epsffile{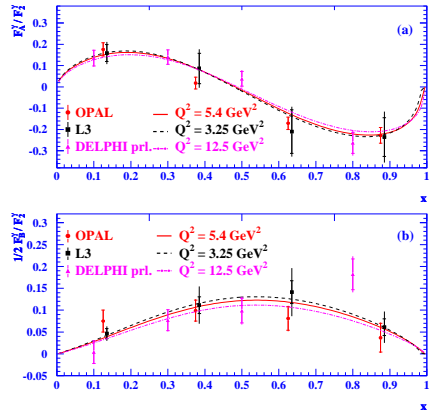}
           }
\end{center}
\caption{The measured ratios
$F_A^{\gamma}/F_2^{\gamma}$ and $1/2\cdot F_B^{\gamma}/F_2^{\gamma}$
compared to the QED prediction~\protect\cite{bib-nisius}.}  
\label{fig1}
\end{figure}
The QED structure function $F_2^{\gamma}$ 
has been measured in the process 
e$^+$e$^-\to$e$^+$e$^-\mu^+\mu^-$. In addition,
the measurement of the distribution of the azimuthal angle $\chi$
between the electron scattering plane and the
plane containing the muon pair in the $\gamma^{\ast}\gamma$ 
centre-of-mass system gives access to the structure 
functions $F_A^{\gamma}$ and $F_B^{\gamma}$.  
They are related to the 
transverse-longitudinal (A) and the transverse-transverse (B) interference
in the interaction of the transverse real photon with the virtual photon.
The LEP measurements~\cite{bib-delphiqed,bib-l3qed,bib-opalqed} 
are shown in Fig.~\ref{fig1}. 
Both structure functions are found to be significantly
different from zero and the ratios are well described 
by QED~\cite{bib-nisius}. 
\section{Hadronic Structure Functions}
The measurement of hadronic structure functions is considerably
more difficult due to the necessity to reconstruct $x$ from
the hadronic final state in the detector (Eq.~\ref{eq2}).
Significant progress has been made recently in reducing
the systematic errors due to unfolding and hadronisation
uncertainties. ALEPH, L3 and OPAL have compared their combined
data to the PHOJET and HERWIG generators~\cite{bib-lepwg}.
An unbiased tune using informations from HERA has improved
HERWIG significantly.

Furthermore new methods for regularised unfolding like the
maximum entropy method or the singular value decomposition
method have been used. ALEPH and OPAL have introduced
two-dimensional unfolding and improved treatment of hadronic
energy in the forward region. L3 is applying energy-momentum
conservation using kinematic information from both hadrons and 
the electrons. 
\begin{figure}[hbtp]
\begin{center}
     \mbox{
         \epsfxsize=160pt
          \epsffile{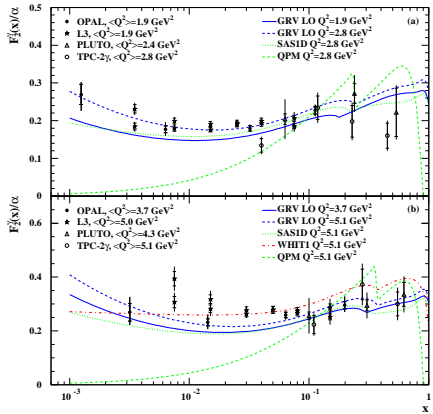}
           }
\end{center}
\caption{The measured hadronic structure function 
$F_2^{\gamma}$ compared to the GRV-LO~\protect\cite{bib-grv}, 
SaS-1D~\protect\cite{bib-sas} 
and the WHIT1~\protect\cite{bib-whit1} parametrisations
and to the QPM model. In the case of L3 the values obtained
using PHOJET and TWOGAM for unfolding are shown separately.}
\label{fig2}
\end{figure}
The uncertainty on the measurements shown in Fig.~\ref{fig2} are
therefore considerably reduced~\cite{bib-l3had,bib-opalhad}. 

The hadron-like component dominates at low $x$ and there could be a first
indication of the low $x$ rise of the photon structure function
expected from QCD evolution.
\begin{figure}[hbtp]
\begin{center}
\hskip -4mm
     \mbox{
         \epsfxsize=195pt
          \epsffile{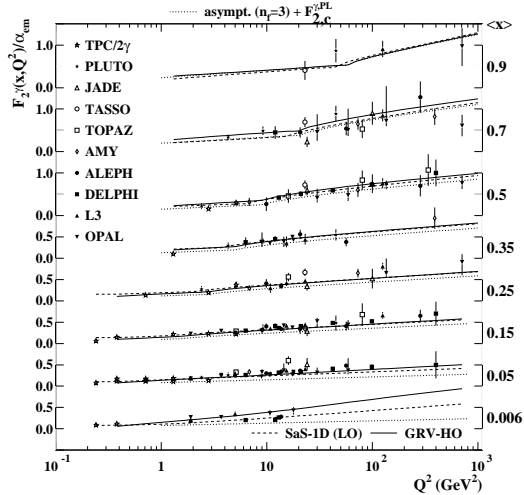}
           }
\end{center}
\caption{The $Q^2$ dependence of the hadronic structure function 
$F_2^{\gamma}$ in bins of $x$
compared to the GRV-HO, SaS-1D and the asymptotic prediction.}
\label{fig3}
\end{figure}

The $Q^2$ dependence of the structure function 
$F_2^{\gamma}$ in bins of $x$ is shown in Fig.~\ref{fig3} for
all currently available measurements. The data are
compared to the GRV-HO and the SaS-1D parametrisation,
and to the sum of the asymptotic prediction~\cite{bib-asy} for 3 light flavours
and the point-like part of the charm structure function taken from GRV.
Positive scaling violation of the photon structure function
is observed in all $x$ ranges - different from the proton - due
to the regular QCD evolution at low $x$ and due to the inhomogeneous term 
($\gamma\to\mbox{q}\overline{\mbox{q}}$) at larger $x$.
As expected, the asymptotic prediction 
fails to describe the data at low $x$, where
the non-perturbative hadron-like contribution dominates, whereas
all models give a reasonable description of the medium to high $x,Q^2$ region.
\section{Charm Structure Function}
In Fig.~\ref{fig3} the charm threshold is clearly visible. 
Above the kinematic threshold for charm production, the c and u
contribution to the point-like part 
of the photon structure function are of similar size. 
OPAL has measured the charm
\begin{figure}[hbtp]
\begin{center}
     \mbox{
         \epsfxsize=190pt
          \epsffile{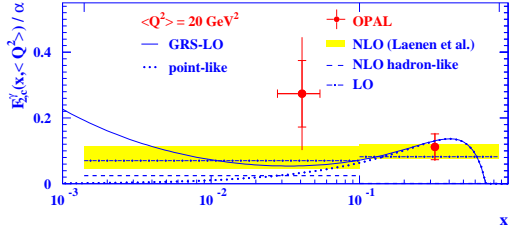}
           }
\end{center}
\caption{Charm structure function $F_{2,\rm c}^{\gamma}$ of the photon
as function of $x$ for $\langle Q^2\rangle=20$~GeV$^2$.}
\label{fig5}
\end{figure}
structure function of the photon for the first time using
D$^{\ast}$ decays~\cite{bib-opalcharm}.
The region $x>0.1$ - which is dominated by the point-like
component - is in good agreement with a NLO calculation~\cite{bib-laenen}.
In the region $x<0.1$ the measurement suggests the existence of
a hadron-like component with currently large errors. These
uncertainties are expected to be significantly reduced in the future
due to higher statistics and better MC modelling of
charm production.

\section{Virtual Photon Structure}
In addition to the structure function of
(quasi-)real photons, i.e. $P^2\approx 0$, the effective structure function
of virtual photons can be measured if $Q^2>>P^2>>\Lambda^2_{\rm QCD}$.
This was first done by PLUTO~\cite{bib-pluto}.
\begin{figure}[hbtp]
\begin{center}
     \mbox{
         \epsfxsize=160pt
          \epsffile{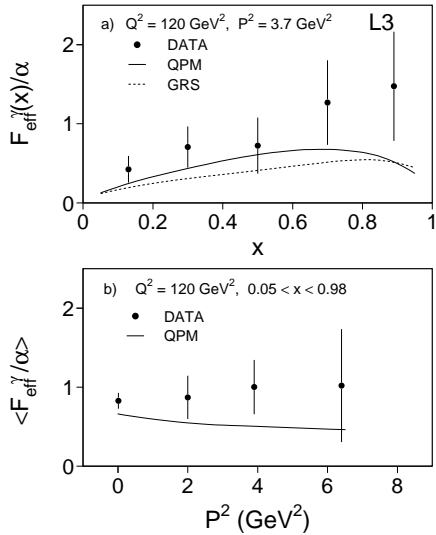}
           }
\end{center}
\caption{Effective structure function of the virtual photon
as function of $x$ and $P^2$.}
\label{fig4}
\end{figure}
For real photons only
the cross-sections $\sigma_{\rm LT}$ and $\sigma_{\rm TT}$
contribute, where the indices refer to the longitudinal and transverse
helicity states of the probe and target photon, respectively,
i.e. $F_2^{\gamma}\simeq \sigma_{\rm LT}+\sigma_{\rm TT}$.
For $P^2>>0$ other helicity states have to be taken into
account, leading to the definition of the effective
structure function
$F_{\rm eff}^{\gamma}\simeq \sigma_{\rm LT}+\sigma_{\rm TT}
+\sigma_{\rm TL}+\sigma_{\rm LL}$ (interference terms are neglected).
This effective structure function measured
by L3~\cite{bib-l3vir} is shown in Fig.~\ref{fig4}.
We expect the hadron-like part of the parton densities at low
$x$ to decrease with increasing virtuality of the photon. 
In Fig.~\ref{fig4}b the QPM approximation of the point-like part
therefore fails to describe the data point at $P^2=0$. The shape of the $P^2$
dependence is consistent with the QPM ansatz but the
errors are still large. Much more precise data is to be expected
from LEP on virtual photon structure in the next years.

\end{document}